\newtheorem{theorem}{Theorem}
\newtheorem{definition}{Definition}

\documentclass[12pt,thmsa]{article}%
\usepackage{amsfonts}
\usepackage{float}
\usepackage{amsmath}%
\usepackage{amssymb}%
\usepackage{graphicx}
\begin{document}

\title{{\LARGE MUTUAL\ ENTROPY\ IN\ QUANTUM\ INFORMATION\ AND\ INFORMATION\ GENETICS}}
\author{Masanori OHYA\\Department of Information Sciences,\\Science University of Tokyo\\Noda City, Chiba 278-8510, JAPAN}
\date{}
\maketitle

\begin{abstract}
After Shannon, entropy becomes a fundamental quantity to describe not only
uncertainity or chaos of a system but also information carried by the system.
Shannon's important discovery is to give a mathematical expression of the
mutual entropy (information), information transmitted from an input system to
an output system, by which communication processes could be analyzed on the
stage of mathematical science. In this paper, first we review the quantum
mutual entropy and discuss its uses in quantum information theory, and
secondly we show how the classical mutual entropy can be used to analyze
genomes, in particular, those of HIV.

\end{abstract}

\section{Introduction}

The study of mutual entropy (information) and capacity in classical system was
extensively done after Shannon by several authors like Kolmogorov \cite{Kol}
and Gelfand \cite{GY}. In quantum systems, there have been several definitions
of the mutual entropy for classical input and quantum output
\cite{BS,Hol,Ing,Lev}. In 1983, the author defined \cite{O2} the fully quantum
mechanical mutual entropy by means of the relative entropy of Umegaki
\cite{U}, and it has been used to compute the capacity of quantum channel for
quantum communication process; quantum input-quantum output \cite{OPW1,OPW2}.

A correlated state in quantum systems, so-called quantum entangled state or
quantum entanglement, are used to study quantum information, in particular,
quantum computation, quantum teleportation, quantum cryptography
\cite{BO,Ben,BBPSSW,Eke,IOS,JB,OPW2,Sch1,Sch2}. Recently Belavkin and Ohya
\cite{BO} characterized the entangled states and introduced the mutual entropy
for entangled states to measure the degree of the entanglement.

In part I of this paper, we mainly discuss the following two topics; (1) the
quantum mutual entropy, the capacity of quantum channel and their uses in
quantum communication; (2)the quantum mutual entropy for entangled states.

Genome sequences is considered to carry information, and the information is
stored in base or amino acid sequences so that it originates the life itself.
In part II of this paper, we present how information theory is used to
investigate the ''information'' stored in DNA. In particular, we shall discuss
the uses of several informations (entropies) and the artificial codes to
analyze genomes of, for instance, HIV.

\part{Quantum Information}

\section{\textbf{Quantum Mutual Entropy}}

The quantum mutual entropy was introduced in \cite{O2} for a quantum input and
quantum output, namely, for purely quantum channel, and it was generalized for
a general quantum system described by C*-algebraic terminology\cite{O4}. We
here review the quantum mutual entropy in usual quantum system described by a
Hilbert space.

Let $\mathcal{H}$ be a Hilbert space for an input space,\textbf{\ }%
$B\mathcal{(H)}$ be the set of all bounded linear operators on $\mathcal{H}$
and $\mathcal{\ S(H)}${\scriptsize \ }be the set of all density operators on
$\mathcal{H}.$ An output space is described by another Hilbert space
$\overset{\sim}{\mathcal{H}}$, but often $\mathcal{H=}\overset{\sim
}{\mathcal{H}}$. A channel from the input system to the output system is a
mapping $\Lambda$* from $\mathcal{S(H)}$ to $\mathcal{S(\overset{\sim
}{\mathcal{H}})}$ \cite{O1}. A channel $\Lambda$* is said to be completely
positive if the dual map $\Lambda$ satisfies the following condition:
$\Sigma_{k,j=1}^{n}$ $A_{k}^{*}\Lambda(B_{k}^{*}B_{j})A_{j}\geq0$ for any
$n\in$\textbf{N} and any $A_{j}\in B(\mathcal{H}),B_{j}\in B(\overset{\sim
}{\mathcal{H}})$.

An input state $\rho$ $\in\mathcal{S(H)}$ is sent to the output system through
a channel $\Lambda$*, so that the output state is written as $\overset{\sim
}{\rho}\equiv\Lambda^{*}\rho.$ Then it is important to ask how much
information of $\rho$ is correctly sent to the output state $\Lambda^{*}\rho.$
This amount of information transmitted from input to output is expressed by
the mutual entropy in Shannon's theory.

In order to define the quantum mutual entropy, we first mention the entropy of
a quantum state introduced by von Neumann\cite{Neu}. For a state $\rho,$ there
exists a unique spectral decomposition
\begin{equation}
\rho=\Sigma_{k}\lambda_{k}P_{k},\label{2.1}%
\end{equation}
where $\lambda_{k}$ is an eigenvalue of $\rho$ and $P_{k}$ is the associated
projection for each $\lambda_{k}$. The projection $P_{k}$ is not
one-dimensional when $\lambda_{k}$ is degenerated, so that the spectral
decomposition can be further decomposed into one-dimensional projections. Such
a decomposition is called a Schatten decomposition, namely,%

\begin{equation}
\rho=\Sigma_{k}\lambda_{k}E_{k},\label{2.2}%
\end{equation}
where $E_{k}$ is the one-dimensional projection associated with $\lambda_{k} $
and the degenerated eigenvalue $\lambda_{k}$ repeats dim$P_{k}$ times. This
Schatten decomposition is not unique unless every eigenvalue is
non-degenerated. Then the entropy (von Neumann entropy) $S\left(  \rho\right)
$ of a state $\rho$ is defined by%

\begin{equation}
S\left(  \rho\right)  =-tr\rho\log\rho,\label{2.3}%
\end{equation}
which equals to the Shannon entropy of the probability distribution $\left\{
\lambda_{k}\right\}  $ :%

\begin{equation}
S\left(  \rho\right)  =-\sum_{k}\lambda_{k}\log\lambda_{k}.\label{2.4}%
\end{equation}

The quantum mutual entropy was introduced on the basis of the above von
Neumann entropy for purely quantum communication processes. The mutual entropy
depends on an input state $\rho$ and a channel $\Lambda^{*}$, so it is denoted
by $I\left(  \rho;\Lambda^{*}\right)  $, which should satisfy the following conditions:

(1) The quantum mutual entropy is well-matched to the von Neumann entropy.
Furthermore, if a channel is trivial, i.e., $\Lambda^{*}=$ identity map, then
the mutual entropy equals to the von Neumann entropy: $I\left(  \rho
;id\right)  $ = $S\left(  \rho\right)  $.

(2) When the system is classical, the quantum mutual entropy reduces to
classical one.

(3) Shannon's fundamental inequality 0$\leq$ $I\left(  \rho;\Lambda
^{*}\right)  \leq S\left(  \rho\right)  $ is held.

Before mentioning the quantum mutual entropy, we briefly review the classical
mutual entropy. Let $\left(  \Omega,\mathcal{F}\right)  $ , $\left(
\overline{\Omega},\overline{\mathcal{F}}\right)  $be an input and output
measurable spaces, respectively, and $P\left(  \Omega\right)  ,$ $P\left(
\overline{\Omega}\right)  $ are the corresponding set of all probability
measures (states). A channel $\Lambda^{*}$ is a mapping from $P\left(
\Omega\right)  $ to $P\left(  \overline{\Omega}\right)  $ and its dual
$\Lambda$ is a map from the set $B\left(  \Omega\right)  $ of all Baire
measurable functions on $\Omega$ to $B\left(  \overline{\Omega}\right)  .$ For
an input state $\mu\in P\left(  \Omega\right)  ,$ the output state
$\overline{\mu}=$ $\Lambda^{*}\mu$ and the joint state (probability measure)
$\Phi$ is given by%

\begin{equation}
\Phi\left(  Q\times\text{ }\overline{Q}\right)  =\int_{\overline{Q}}%
\Lambda\left(  1_{Q}\right)  d\mu,\text{ }Q\in\mathcal{F},\text{ }\overline
{Q}\in\overline{\mathcal{F}},
\end{equation}
where 1$_{Q}$ is the characteristic function on $\Omega:$ 1$_{Q}\left(
\omega\right)  =\left\{
\begin{array}
[c]{ll}%
1 & \left(  \omega\in Q\right) \\
0 & \left(  \omega\notin Q\right)
\end{array}
\right.  .$ The classical entropy, relative entropy and mutual entropy are
defined as follows:%

\begin{equation}
S\left(  \mu\right)  =\sup\left\{  -\sum_{k=1}^{n}\mu\left(  A_{k}\right)
\log\mu\left(  A_{k}\right)  ;\left\{  A_{k}\right\}  \in\mathcal{P}\left(
\Omega\right)  \right\}  ,
\end{equation}

\begin{equation}
S\left(  \mu,\nu\right)  =\sup\left\{  \sum_{k=1}^{n}\mu\left(  A_{k}\right)
\log\frac{\mu\left(  A_{k}\right)  }{\nu\left(  A_{k}\right)  };\left\{
A_{k}\right\}  \in\mathcal{P}\left(  \Omega\right)  \right\}  ,
\end{equation}

\begin{equation}
I\left(  \mu;\Lambda^{*}\right)  =S\left(  \Phi,\mu\otimes\Lambda^{*}%
\mu\right)  ,
\end{equation}
where $\mathcal{P}\left(  \Omega\right)  $ is the set of all finite partitions
on $\Omega,$ that is, $\left\{  A_{k}\right\}  \in\mathcal{P}\left(
\Omega\right)  $ iff $A_{k}$ $\in\mathcal{F}$ with $A_{k}\cap A_{j}=\emptyset$
$\left(  k\neq j\right)  $and $\cup_{k=1}^{n}A_{k}=$ $\Omega.$

In order to define the quantum mutual entropy, we need the joint state (it is
called ''compound state'' in the sequel) describing the correlation between an
input state $\rho$ and the output state $\Lambda^{*}\rho$ and the quantum
relative entropy. A finite partition of $\Omega$ in classical case corresponds
to an orthogonal decomposition $\left\{  E_{k}\right\}  $ of the identity
operator I of $\mathcal{H}$ in quantum case because the set of all orthogonal
projections is considered to make an event system for a quantum system. It is
known \cite{OP}that the following equality holds%

\[
\sup\left\{  -\sum_{k}tr\rho E_{k}\log tr\rho E_{k};\left\{  E_{k}\right\}
\right\}  =-tr\rho\log\rho,
\]
and the supremum is attained when $\left\{  E_{k}\right\}  $ is composed of
the Schatten decomposition of $\rho.$ Therefore the Schatten decomposition is
used to define the compound state and the quantum mutual entropy.

The compound state $\theta_{E}$ (corresponding to joint state in CS) of $\rho$
and $\Lambda^{*}\rho$ was introduced in \cite{O2,O3}, which is given by%

\begin{equation}
\theta_{E}=\sum_{k}\lambda_{k}E_{k}\otimes\Lambda^{*}E_{k},\label{2.5}%
\end{equation}
where $E$ stands for a Schatten decomposition of $\rho,$ so that the compound
state depends on how we decompose the state $\rho$ into basic states
(elementary events), in other words, how to see the input state.

The relative entropy for two states $\rho$ and $\sigma$ is defined by Umegaki
\cite{U} and Lindblad \cite{Lin}, which is written as%

\begin{equation}
S\left(  \rho,\sigma\right)  =\left\{
\begin{array}
[c]{ll}%
tr\rho\left(  \log\rho-\log\sigma\right)  & \left(  \text{when }\overline{
ran\rho}\subset\overline{ran\sigma}\right) \\
\infty & \left(  \text{otherwise}\right)
\end{array}
\right.
\end{equation}

Then we can define the mutual entropy by means of the compound state and the
relative entropy \cite{O2}, that is,%

\begin{equation}
I\left(  \rho;\Lambda^{*}\right)  =\sup\left\{  S\left(  \theta_{E}%
,\rho\otimes\Lambda^{*}\rho\right)  ;E=\left\{  E_{k}\right\}  \right\}  ,
\end{equation}
where the supremum is taken over all Schatten decompositions. Some
computations reduce it to the following form:%

\begin{equation}
I\left(  \rho;\Lambda^{*}\right)  =\sup\left\{  \sum_{k}\lambda_{k}S\left(
\Lambda^{*}E_{k},\Lambda^{*}\rho\right)  ;E=\left\{  E_{k}\right\}  \right\}
,
\end{equation}
This mutual entropy satisfies all conditions (1)$\sim$(3) mentioned above.

When the input system is classical, an input state $\rho$ is given by a
probability distribution or a probability measure, in either case, the
Schatten decomposition of $\rho$ is unique, namely, for the case of
probability distribution ; $\rho=\left\{  \lambda_{k}\right\}  ,$%

\begin{equation}
\rho=\sum_{k}\lambda_{k}\delta_{k},
\end{equation}
where $\delta_{k}$ is the delta measure, that is,%

\begin{equation}
\delta_{k}\left(  j\right)  =\delta_{k,j}=\{_{0(k\neq j)}^{1(k=j)},\forall j.
\end{equation}
Therefore for any channel $\Lambda^{*},$ the mutual entropy becomes%

\begin{equation}
I\left(  \rho;\Lambda^{*}\right)  =\sum_{k}\lambda_{k}S\left(  \Lambda
^{*}\delta_{k},\Lambda^{*}\rho\right)  ,
\end{equation}
which equals to the following usual expression of Shannon when it is well-defined:%

\begin{equation}
I\left(  \rho;\Lambda^{*}\right)  =S\left(  \Lambda^{*}\rho\right)  -\sum
_{k}\lambda_{k}S\left(  \Lambda^{*}\delta_{k}\right)  ,
\end{equation}
which has been taken as the definition of the mutual entropy for a
classical-quantum(-classical) channel \cite{B2,BS,Hol,Ing,Lev}.

\smallskip Note that the above definition of the mutual entropy (2.12) is
written as%

\begin{align*}
& I\left(  \rho;\Lambda^{*}\right) \\
& =\sup\left\{  \sum_{k}\lambda_{k}S\left(  \Lambda^{*}\rho_{k},\Lambda
^{*}\rho\right)  ;\rho=\sum_{k}\lambda_{k}\rho_{k}\in F_{o}\left(
\rho\right)  \right\}  ,
\end{align*}
where $F_{o}\left(  \rho\right)  $ is the set of all orthogonal finite
decompositions of $\rho$ \cite{O11}$.$

More general formulation of the mutual entropy for general quantum systems was
done \cite{O4,IKO} in C*dynamical system by using Araki's or Uhlmann's
relative entropy\cite{Ara,Uhl,OP}. This general mutual entropy contains all
other cases including measure theoretic definition of Gelfand and Yaglom
\cite{GY}.

\section{Communication Processes}

The information communication process is mathematically set as follows: M
messages are sent to a receiver and the $k$th message $\omega^{\left(
k\right)  }$ occurs with the probability $\lambda_{k}.$ Then the occurrence
probability of each message in the sequence $\left(  \omega^{\left(  1\right)
},\omega^{\left(  2\right)  },\cdot\cdot\cdot,\omega^{\left(  M\right)
}\right)  $of M messages is denoted by $\rho=\left\{  \lambda_{k}\right\}  ,$
which is a state in a classical system. If $\xi$ is a classical coding, then
$\xi\left(  \omega\right)  $ is a classical object such as an electric pulse.
If $\xi$ is a quantum coding, then $\xi\left(  \omega\right)  $ is a quantum
object (state) such as a coherent state. Here we consider such a quantum
coding, that is, $\xi\left(  \omega^{\left(  k\right)  }\right)  $ is a
quantum state, and we denote $\xi\left(  \omega^{\left(  k\right)  }\right)  $
by $\sigma_{k}.$ Thus the coded state for the sequence $\left(  \omega
^{\left(  1\right)  },\omega^{\left(  2\right)  },\cdot\cdot\cdot
,\omega^{\left(  M\right)  }\right)  $ is written as%

\begin{equation}
\sigma=\sum_{k}\lambda_{k}\sigma_{k}.
\end{equation}
This state is transmitted through a channel $\gamma$, which is expressed by a
completely positive mapping $\Gamma^{*}$ from the state space of $X$ to that
of $\overset{\sim}{X}$ , hence the output coded quantum state $\overset{\sim
}{\sigma}$ is $\Gamma^{*}\sigma.$ Since the information transmission process
can be understood as a process of state (probability) change, when $\Omega$
and $\overset{\sim}{\Omega}$ are classical and $X$ and $\overset{\sim}{X}$ are
quantum, the process (3.1) is written as%

\begin{equation}
P\left(  \Omega\right)  \overset{\Xi^{*}}{\longrightarrow}\mathcal{S} \left(
\mathcal{H}\right)  \overset{\Gamma^{*}}{\longrightarrow}\mathcal{S(}%
\overset{\sim}{\mathcal{H}})\overset{\overset{\sim}{\Xi}^{*}}{ \longrightarrow
}P(\overset{\sim}{\Omega}),
\end{equation}
where $\Xi^{*}$ $($resp.$\overset{\sim}{\Xi}^{*})$ is the channel
corresponding to the coding $\xi$ (resp.$\overset{\sim}{\xi}$ ) and
$\mathcal{S}\left(  \mathcal{H}\right)  $ (resp.$\mathcal{S(}\overset{\sim
}{\mathcal{H}})$ $)$ is the set of all density operators (states) on
$\mathcal{H}$ (resp.$\overset{\sim}{\mathcal{H}}$ $)$.

We have to be care to study the objects in the above transmission process
(3.1) or (3.3). Namely, we have to make clear which object is going to study.
For instance, if we want to know the information capacity of a quantum channel
$\gamma(=\Gamma^{*}),$ then we have to take $X$ so as to describe a quantum
system like a Hilbert space and we need to start the study from a quantum
state in quantum space $X\ $not from a classical state associated to a
message. If we like to know the capacity of the whole process including a
coding and a decoding, which means the capacity of a channel $\overset{\sim
}{\xi}\circ\gamma\circ\xi(=\overset{\sim}{\Xi}^{*}\circ\ \Gamma^{*}\circ
\Xi^{*})$, then we have to start from a classical state$.$ In any case, when
we concern the capacity of channel, we have only to take the supremum of the
mutual entropy $I\left(  \rho;\Lambda^{*}\right)  $ over a quantum or
classical state $\rho$ in a proper set determined by what we like to study
with a channel $\Lambda^{*}.$ We explain this more precisely in the next section.

\section{Channel Capacity}

We discuss two types of channel capacity in communication processes, namely,
the capacity of a quantum channel $\Gamma^{*}$ and that of a classical
(classical-quantum-classical) channel $\overset{\sim}{\Xi}^{*}\circ
\ \Gamma^{*}\circ\Xi^{*}.$

(1) \textit{Capacity of quantum channel:} The capacity of a quantum channel is
the ability of information transmission of a quantum channel itself, so that
it does not depend on how to code a message being treated as classical object
and we have to start from an arbitrary quantum state and find the supremum of
the quantum mutual entropy. One often makes a mistake in this point. For
example, one starts from the coding of a message and compute the supremum of
the mutual entropy and he says that the supremum is the capacity of a quantum
channel, which is not correct. Even when his coding is a quantum coding and he
sends the coded message to a receiver through a quantum channel, if he starts
from a classical state, then his capacity is not the capacity of the quantum
channel itself. In his case, usual Shannon's theory is applied because he can
easily compute the conditional distribution by a usual (classical) way. His
supremum is the capacity of a classical-quantum-classical channel, and it is
in the second category discussed below.

The capacity of a quantum channel $\Gamma^{*}$ is defined as follows: Let
$\mathcal{S}_{0}(\subset$ $\mathcal{S(H))}$ be the set of all states prepared
for expression of information. Then the capacity of the channel $\Gamma^{*}$
with respect to $\mathcal{S}_{0}$ is defined by%

\begin{equation}
C^{\mathcal{S}_{0}}\left(  \Gamma^{*}\right)  =\sup\{I\left(  \rho;\Gamma
^{*}\right)  ;\rho\in\mathcal{S}_{0}\}.
\end{equation}
Here $I\left(  \rho;\Gamma^{*}\right)  $ is the mutual entropy given in (2.11)
or (2.12) with $\Lambda^{*}=\Gamma^{*}.$ When $\mathcal{S}_{0}=\mathcal{S(H)}$
, $C^{\mathcal{S}(\mathcal{H)}}\left(  \Gamma^{*}\right)  $ is denoted by
$C\left(  \Gamma^{*}\right)  $ for simplicity. The capacity $C\left(
\Gamma^{*}\right)  $ is written as%

\begin{equation}
C\left(  \Gamma^{*}\right)  =\sup\{I\left(  \rho;\Gamma^{*}\right)  ;\rho
\in\mathcal{S}\left(  \mathcal{H}\right)  \},
\end{equation}
where the supremum is taken over all states $\rho$ with its orthogonal pure
decomposition $\sum_{k}\lambda_{k}\rho_{k}$ of $\rho.$ In \cite{OPW1,Mur-O1},
we also considered the pseudo-quantum capacity $C_{p}\left(  \Gamma
^{*}\right)  $ defined by (4.1) with the pseudo-mutual entropy $I_{p}\left(
\rho;\Gamma^{*}\right)  $ where the supremum is taken over all finite
decompositions instead of all orthogonal pure decompositions:
\begin{align}
I_{p}\left(  \rho;\Gamma^{*}\right)   & =\sup\left\{  \sum_{k}\lambda
_{k}S\left(  \Gamma^{*}\rho_{k},\Gamma^{*}\rho\right)  ;\rho=\sum_{k}%
\lambda_{k}\rho_{k},\right. \nonumber\\
& \text{\qquad}\left.  \text{ finite decomposition}\right\}  .
\end{align}
However the pseudo-mutual entropy is not well-matched to the conditions
explained in Sec.2, and it is difficult to be computed numerically. The
relation between $C\left(  \Gamma^{*}\right)  $ and $C_{p}\left(  \Gamma
^{*}\right)  $ was discussed in\cite{OPW1}. From the monotonicity of the
mutual entropy\cite{OP}, we have%

\[
0\leq C^{\mathcal{S}_{0}}\left(  \Gamma^{*}\right)  \leq C_{p}^{\mathcal{S}
_{0}}\left(  \Gamma^{*}\right)  \leq\sup\left\{  S(\rho);\rho\in
\mathcal{S}_{0}\right\}  .
\]

(2) \textit{Capacity of classical-quantum-classical channel:} The capacity of
C-Q-C channel $\overset{\sim}{\Xi}^{*}\circ\ \Gamma^{*}\circ\Xi^{*} $ is the
capacity of the information transmission process starting from the coding of
messages, therefore it can be considered as the capacity including a coding
(and a decoding). As is discussed in Sec.3, an input state $\rho$ is the
probability distribution $\left\{  \lambda_{k}\right\}  $ of messages, and its
Schatten decomposition is unique as (2.9), so the mutual entropy is written by (2.11):%

\begin{align}
& I\left(  \rho;\overset{\sim}{\Xi}^{*}\circ\ \Gamma^{*}\circ\Xi^{*}\right)
\nonumber\\
& =\sum_{k}\lambda_{k}S\left(  \overset{\sim}{\Xi}^{*}\circ\ \Gamma^{*}%
\circ\Xi^{*}\delta_{k},\overset{\sim}{\Xi}^{*}\circ\ \Gamma^{*}\circ\Xi
^{*}\rho\right)  .
\end{align}
If the coding $\Xi^{*}$ is a quantum coding, then $\Xi^{*}\delta_{k}$ is
expressed by a quantum state. Let denote the coded quantum state by
$\sigma_{k}$ and put $\sigma=\Xi^{*}\rho=\sum_{k}\lambda_{k}\sigma_{k}.$ Then
the above mutual entropy is written as%

\begin{equation}
I\left(  \rho;\overset{\sim}{\Xi}^{*}\circ\ \Gamma^{*}\circ\Xi^{*}\right)
=\sum_{k}\lambda_{k}S\left(  \overset{\sim}{\Xi}^{*}\circ\ \Gamma^{*}%
\sigma_{k},\overset{\sim}{\Xi}^{*}\circ\ \Gamma^{*}\sigma\right)  .
\end{equation}
This is the expression of the mutual entropy of the whole information
transmission process starting from a coding of classical messages. Hence the
capacity of C-Q-C channel is%

\begin{equation}
C^{P_{0}}\left(  \overset{\sim}{\Xi}^{*}\circ\ \Gamma^{*}\circ\Xi^{*}\right)
=\sup\{I\left(  \rho;\overset{\sim}{\Xi}^{*}\circ\ \Gamma^{*}\circ\Xi
^{*}\right)  ;\rho\in P_{0}\},
\end{equation}
where $P_{0}(\subset P(\Omega))$ is the set of all probability distributions
prepared for input (a-priori) states (distributions or probability measures).
Moreover the capacity for coding free is found by taking the supremum of the
mutual entropy (4.4) over all probability distributions and all codings
$\Xi^{*}$:%

\begin{equation}
C_{c}^{P_{0}}\left(  \overset{\sim}{\Xi}^{*}\circ\ \Gamma^{*}\right)
=\sup\{I\left(  \rho;\overset{\sim}{\Xi}^{*}\circ\ \Gamma^{*}\circ\Xi
^{*}\right)  ;\rho\in P_{0},\Xi^{*}\}.
\end{equation}
The last capacity is for both coding and decoding free and it is given by%

\begin{equation}
C_{cd}^{P_{0}}\left(  \ \Gamma^{*}\right)  =\sup\{I\left(  \rho;\overset{\sim
}{\Xi}^{*}\circ\ \Gamma^{*}\circ\Xi^{*}\right)  ;\rho\in P_{0},\Xi
^{*},\overset{\sim}{\Xi}^{*}\}.
\end{equation}
These capacities $C_{c}^{P_{0}},$ $C_{cd}^{P_{0}}$ do not measure the ability
of the quantum channel $\Gamma^{*}$ itself, but measure the ability of
$\Gamma^{*}$ through the coding and decoding.

Remark that $\sum_{k}\lambda_{k}S(\Gamma^{*}\sigma_{k})$ is finite, then (4.4) becomes%

\begin{equation}
I\left(  \rho;\overset{\sim}{\Xi}^{*}\circ\ \Gamma^{*}\circ\Xi^{*}\right)
=S(\overset{\sim}{\Xi}^{*}\circ\Gamma^{*}\sigma)-\sum_{k}\lambda_{k}%
S(\overset{\sim}{\Xi}^{*}\circ\Gamma^{*}\sigma_{k}).
\end{equation}
Further, if $\rho$ is a probability measure having a density function
$f(\lambda)$ and each $\lambda$ corresponds to a quantum coded state
$\sigma(\lambda),$ then $\sigma=\int f(\lambda)$ $\sigma(\lambda)d\lambda$ and%

\begin{align}
& I\left(  \rho;\overset{\sim}{\Xi}^{*}\circ\ \Gamma^{*}\circ\Xi^{*}\right)
\nonumber\\
& =S(\overset{\sim}{\Xi}^{*}\circ\Gamma^{*}\sigma)-\int f(\lambda
)S(\overset{\sim}{\Xi}^{*}\circ\Gamma^{*}\sigma(\lambda))d\lambda.
\end{align}
This is bounded by%

\[
S(\Gamma^{*}\sigma)-\int f(\lambda)S(\Gamma^{*}\sigma(\lambda))d\lambda,
\]
which is called the Holevo bound and is computed in several occasions
\cite{YO,OPW1}

The above three capacities $C^{P_{0}},$ $C_{c}^{P_{0}},$ $C_{cd}^{P_{0}}$
satisfy the following inequalities
\begin{align*}
0  & \leq C^{P_{0}}\left(  \overset{\sim}{\Xi}^{*}\circ\ \Gamma^{*}\circ
\Xi^{*}\right)  \leq C_{c}^{P_{0}}\left(  \overset{\sim}{\Xi}^{*}\circ
\ \Gamma^{*}\right) \\
& \leq C_{cd}^{P_{0}}\left(  \ \Gamma^{*}\right)  \leq\sup\left\{
S(\rho);\rho\in P_{o}\right\}
\end{align*}
where $S(\rho)$ is not the von Neumann entropy but the Shannon entropy:
-$\sum\lambda_{k}\log\lambda_{k}.$

The capacities (4.1), (4.6),(4.7) and (4.8) are generally different. Some
misunderstandings occur due to forgetting which channel is considered. That
is, we have to make clear what kind of the ability (capacity) is considered,
the capacity of a quantum channel itself or that of a
classical-quantum(-classical ) channel. The computation of the capacity of a
quantum channel was carried in several models in \cite{OPW1,OPW2}

\section{Quantum Entanglements}

Recently the quantum entangled state has been mathematically studied
\cite{BBPSSW,Maj,Sch1}, in which the entangled state is defined by a state not
written as a form $\sum_{k}\lambda_{k}\rho_{k}\otimes\sigma_{k}$ with any
states $\rho_{k}$ and $\sigma_{k}.$ A state written as above is called a
separable state, so that an entangled state is a state not belonged to the set
of all separable states. However it is obvious that there exist several
correlated states written as separable forms. Such correlated states have been
discussed in several contexts in quantum probability such as quantum filtering
\cite{B2}, quantum compound state \cite{O2}, quantum Markov state \cite{A} and
quantum lifting \cite{AO}. In \cite{BO}, we showed a mathematical construction
of quantum entangled states and gave a finer classification of quantum sates.

For the (separable) Hilbert space $\mathcal{K}$ of a quantum system, let
$\mathcal{A\equiv}$ $B\left(  \mathcal{K}\right)  $ be the set of all linear
bounded operators {on }$\mathcal{K}$. A normal state $\varphi$ on
$\mathcal{\ \ A}$ can be expressed as $\varphi\left(  A\right)
=tr_{\mathcal{G}}\kappa^{\dagger}A\kappa,$ $A\in\mathcal{A}$, where
$\mathcal{G}$ is another separable Hilbert space, $\kappa$ is a linear
Hilbert-Schmidt operator from $\mathcal{G}$ to $\mathcal{K}$ and
$\kappa^{\dagger}$ is the adjoint operator of $\kappa$ from $\mathcal{K}$ to
$\mathcal{G}$. The (unique) density operator $\sigma\in\mathcal{A}$ associated
to the state $\varphi:\varphi\left(  A\right)  =trA\sigma$, $A\in\mathcal{A}$,
is written by $\kappa$ such as $\sigma=\kappa\kappa^{\dagger}.$ This $\kappa$
is called the amplitude operator, and it is called just the amplitude if
$\mathcal{G}$ is one dimensional space $\mathbb{C}$, corresponding to the pure
state $\varphi\left(  A\right)  =\kappa^{\dagger}A\kappa$ for a $\kappa
\in\mathcal{K}$ with $\kappa^{\dagger}\kappa=\Vert\kappa\Vert^{2}=1$. In
general, $\mathcal{G}$ is not one dimensional, the dimensionality
$\dim\mathcal{G}$ must be not less than $\dim\sigma\mathcal{K}.$

Since $\mathcal{G}$ is separable, $\mathcal{G}$ is realized as a subspace of
$l^{2}(\mathbf{N})$ of complex sequences $(i.e.,\zeta^{\bullet}=\left(
\zeta^{n}\right)  ,$ $\zeta^{n}\in\mathbb{C}$, $n\in\mathbf{N}$ with
$\sum\left|  \zeta^{n}\right|  ^{2}<+\infty)$, so that any vector
$\zeta^{\bullet}=(\zeta^{n}$) represents a vector $\zeta=\sum\zeta
^{n}|n\rangle$ in the standard basis $\left\{  |n\rangle\right\}
\in\mathcal{G}$ of $l^{2}(\mathbf{N})$ .

Given the amplitude operator $\kappa$, one can define not only the states
$\sigma\equiv\kappa\kappa^{\dagger}$ and $\rho\equiv$ $\kappa^{\dagger}\kappa$
on the algebras $\mathcal{A}\left(  =B\left(  \mathcal{K} \right)  \right)  $
and $\mathcal{B}\left(  =B\left(  \mathcal{G}\right)  \right)  $ but also an
entanglement state $\Theta$ on the algebra $\mathcal{B}\otimes\mathcal{A}$ of
all bounded operators on the tensor product Hilbert space $\mathcal{G}%
\otimes\mathcal{K}$ by%

\[
\Theta\left(  B\otimes A\right)  =tr_{\mathcal{G}}B\kappa^{\dagger}%
A\kappa=tr_{\mathcal{K}}A\kappa B\kappa^{\dagger}
\]
for any $B\in\mathcal{B}$. This state is pure as it is the case of
$\mathcal{F}=\mathbb{C}$ in the theorem below, and it satisfies the marginal
conditions: For any $B\in\mathcal{B},A\in\mathcal{A}$,%

\[
\Theta\left(  B\otimes I\right)  =tr_{\mathcal{G}}B\rho,\quad\Theta\left(
I\otimes A\right)  =tr_{\mathcal{K}}A\sigma.\quad
\]

\begin{theorem}
\cite{BO}Let $\Theta:\mathcal{B}\otimes\mathcal{A}\rightarrow\mathbb{C}$ be a
state
\begin{equation}
\Theta\left(  B\otimes A\right)  =tr_{\mathcal{F}}\psi^{\dagger}\left(
B\otimes A\right)  \psi,
\end{equation}
defined by an amplitude operator $\psi$ on a separable Hilbert space
$\mathcal{E}$ into the tensor product Hilbert space $\mathcal{G}%
\otimes\mathcal{K}$ ; $\psi:\mathcal{E}\rightarrow\mathcal{G}\otimes
\mathcal{K}$ with $tr_{\mathcal{F}}\psi^{\dagger}\psi=1$. Then there exists an
amplitude operator $\kappa:\mathcal{G}\rightarrow\mathcal{F}\otimes
\mathcal{K}$ such that the state $\Theta$ can be achieved by an entanglement
\begin{equation}
\Theta\left(  B\otimes A\right)  =tr_{\mathcal{G}}B\kappa^{\dagger}\left(
I\otimes A\right)  \kappa=tr_{\mathcal{F}\otimes\mathcal{K}}\left(  I\otimes
A\right)  \kappa B\kappa^{\dagger}%
\end{equation}
The entangling operator $\kappa$ is uniquely defined up to a unitary
transformation of the minimal space $\mathcal{F}$.
\end{theorem}

The entangled state (5.2) is written as
\begin{equation}
\Theta\left(  B\otimes A\right)  =tr_{\mathcal{G}}B\phi\left(  A\right)
=tr_{\mathcal{K}}A\phi_{*}\left(  B\right)  ,
\end{equation}
where $\phi\left(  A\right)  \equiv\kappa^{\dagger}\left(  I\otimes A\right)
\kappa$ is in the predual space $\mathcal{B}_{*}\subset\mathcal{B} $ of all
trace-class operators in $\mathcal{G}$, and $\phi_{*}\left(  B\right)  \equiv
tr_{\mathcal{F}}\kappa B\kappa^{\dagger}$ is in $\mathcal{A}_{*}%
\subset\mathcal{A}$. The map $\phi$ is the Steinspring form of the general
completely positive map $\mathcal{A}\rightarrow\mathcal{B}_{*}$, written in
the eigen-basis $\left\{  \left|  n\right\rangle \right\}  $ of $\mathcal{G}%
\subseteq l^{2}\left(  \mathbb{N}\right)  $ of the density operator $\rho
=\phi\left(  I\right)  $ as
\begin{equation}
\phi\left(  A\right)  =\sum_{m,n}|m\rangle\kappa_{m}^{\dagger}\left(  I\otimes
A\right)  \kappa_{n}\langle n|,\quad A\in\mathcal{A}\label{*3}%
\end{equation}
where $\kappa_{n}$ is the vector in $\mathcal{F}\otimes\mathcal{K}$ such that
$\kappa=\sum_{n}\kappa_{n}\langle n|$. The dual operation $\phi_{*}$ is the
Kraus form of the general completely positive map $\mathcal{B}\rightarrow
\mathcal{A}_{*}$, given in this basis as
\begin{equation}
\phi_{*}\left(  B\right)  =\sum_{n,m}\left\langle n\right|  B\left|
m\right\rangle tr_{\mathcal{F}}\kappa_{n}\kappa_{m}^{\dagger},\quad
B\in\mathcal{B}.\label{*4}%
\end{equation}
It corresponds to the general form of the density operator
\begin{equation}
\theta_{\phi}=\sum_{m,n}|n\rangle\langle m|\otimes tr_{\mathcal{F}}\kappa
_{n}\kappa_{m}^{\dagger}\label{*5}%
\end{equation}
for the entangled state $\Theta$ with the weak orthogonality property
\begin{equation}
tr_{\mathcal{F}\otimes\mathcal{K}}\kappa_{n}\kappa_{m}^{\dagger}=p_{n}%
\delta_{n}^{m}=\kappa_{m}^{\dagger}\kappa_{n}.
\end{equation}

\begin{definition}
The dual map $\phi_{*}:\mathcal{B}\rightarrow\mathcal{A}_{*}$ to a completely
positive map $\phi:\mathcal{A}\rightarrow\mathcal{B}_{*}$, normalized as
$tr_{\mathcal{G}}\phi\left(  I\right)  =1$, is called the quantum entanglement
of the state $\rho=\phi\left(  I\right)  $ on $\mathcal{\ B}$ to the state
$\sigma=\phi_{*}\left(  I\right)  $ on $\mathcal{A}$. The entanglement by
$\phi\left(  A\right)  =\sigma^{1/2}A\sigma^{1/2}$ of the state $\rho=\sigma$
on the algebra $\mathcal{B}=\mathcal{A}$ given by the standard entangling
operator $\kappa=\sigma^{1/2}$ is called standard.
\end{definition}

A compound state, playing the similar role as the joint input-output
probability measures in classical systems, was introduced in \cite{O2} as
explained in Sec.2. It corresponds to a particular diagonal type
\[
\theta_{\phi}=\sum_{n}|n\rangle\langle n|\otimes tr_{\mathcal{F}}\kappa
_{n}\kappa_{n}^{\dagger}
\]
of the entangling map (\ref{*4}) in the eigen-basis(Schatten decomposition) of
the density operator $\rho=\sum p_{n}|n\rangle\langle n|$. Therefore the
entangled states, generalizing the compound state, also play the role of the
joint probability measures.

The diagonal entanglements can be considered as a quantum correspondences of
symbols $\left\{  1,\cdots,n,\cdots\right\}  $ to quantum states. The general
entangled states $\Theta$ are described by the density operators $\theta
_{\phi}$ of the form (\ref{*5}) which is not necessarily diagonal in the
eigen-representation of the density operator $\rho=\sum_{n}p_{n}%
|n\rangle\langle n|$. Such nondiagonal entangled states were called in
\cite{O4} the quasicompound (q-compound) states, so we can call also the
nondiagonal entanglement the quantum quasi-correspondence (q-correspondence)
in contrast to the d-correspondences, described by the diagonal entanglements,
giving rise to the d-compound states.

Take $tr_{\mathcal{F}}\kappa_{n}\kappa_{n}^{\dagger}\equiv\upsilon_{n}%
\upsilon_{n}^{\dagger},$ $\upsilon_{n}\in\mathcal{K}$. The density operator
\begin{equation}
\theta=\sum_{n}|n\rangle\langle n|\otimes\sigma_{n},\quad\sigma_{n}%
=p_{n}\upsilon_{n}\upsilon_{n}^{\dagger}\label{6.1}%
\end{equation}
define the compound states on $\mathcal{B}\otimes\mathcal{A}$, giving the
quantum correspondences $n\mapsto|n\rangle\langle n|$ with the probabilities
$p_{n}$. The entanglement with (\ref{6.1}) is a diagonal entanglement such as
\begin{equation}
\phi_{*}\left(  B\right)  =\sum_{n}p_{n}\langle n|B|n\rangle\upsilon
_{n}\upsilon_{n}^{\dagger}\label{6.2}%
\end{equation}
whose dual is
\begin{equation}
\phi\left(  A\right)  =\sum_{n}p_{n}|n\rangle\upsilon_{n}^{\dagger}%
A\upsilon_{n}\langle n|.
\end{equation}
\newline These entanglements has the stronger orthogonality
\begin{equation}
tr_{\mathcal{F}}\kappa_{n}\kappa_{m}^{\dagger}=p_{n}\upsilon_{n}\upsilon
_{n}^{\dagger}\delta_{n}^{m},\label{6.4}%
\end{equation}
for the amplitudes $\kappa_{n}\in\mathcal{F}\otimes\mathcal{K}$ of the
decomposition $\kappa=\sum_{n}\kappa_{n}\langle n|$ in comparison with the
weak orthogonality of $\kappa_{n}$ in (\ref{*5}).

\begin{definition}
The positive diagonal map
\begin{equation}
\phi_{*}\left(  B\right)  =\sum_{n}\langle n|B|n\rangle\sigma_{n}%
\end{equation}
into the subspace of trace-class operation $\mathcal{K}$ with $tr_{\mathcal{G}%
}\phi_{*}\left(  I\right)  =1$, is called quantum d-entanglement with the
input probabilities $p_{n}=tr_{\mathcal{K}}\sigma_{n}$ and the output states
$\omega_{n}=p_{n}^{-1}\sigma_{n}$, and the corresponding compound state
(\ref{2.5}) is called d-compound state. The d-entanglement is called
c-entanglement and compound state is called c-compound if all density
operators $\sigma_{n}$ commute: $\sigma_{m}\sigma_{n}=\sigma_{n}\sigma_{m}$
for all $m$ and $n$.
\end{definition}

Note that due to the commutativity of the operators $B\otimes I$ with
$I\otimes A$ on $\mathcal{G}\otimes\mathcal{K}$, one can treat the
correspondences as the nondemolition measurements in $\mathcal{B}$ with
respect to $\mathcal{A}$. So, the compound state is the state prepared for
such measurements on the input $\mathcal{G}$. It coincides with the mixture of
the states, corresponding to those after the measurement without reading the
sent message. The set of all d-entanglements corresponding to a given Schatten
decomposition of the input state $\rho$ on $\mathcal{A}$ is obviously convex
with the extreme points given by the pure elementary output states $\omega
_{n}$ on $\mathcal{A}$, corresponding to a not necessarily orthogonal
decompositions $\sigma=\sum_{n}\sigma_{n}$ into one-dimensional density
operators $\sigma_{n}=p_{n}\omega_{n}.$

The orthogonal Schatten decompositions $\sigma=\sum_{n}p_{n}\omega_{n}$
correspond to the extreme points of c-entanglements which also form a convex
set with mixed commuting $\omega_{n}$ for a given Schatten decomposition of
$\sigma$. The orthogonal c-entanglements were used in \cite{AO} to construct a
particular type of Accardi's transition expectations \cite{A} and to define
the entropy in a quantum dynamical system via such transition
expectations\cite{BO}.

Thus we classified the entangled states into three categories, namely,
q-entangled state, d-entangled state and c-entangled state, and their rigorous
expressions were given.

\section{Mutual Entropy via Entanglements}

Let us consider the entangled mutual entropy by means of the above three types
compound states. We denote the quantum mutual entropy of the compound state
$\Theta$ achieved by an entanglement $\phi_{*}:$ $\mathcal{B}\rightarrow
\mathcal{A}_{*}$ with the marginals
\begin{equation}
\Theta\left(  B\otimes I\right)  =tr_{\mathcal{G}}B\rho,\;\Theta\left(
I\otimes A\right)  =tr_{\mathcal{K}}A\sigma\label{7.1}%
\end{equation}
by $I_{\phi}\left(  \rho,\sigma\right)  $ or $I_{\phi}\left(  \mathcal{A}%
,\mathcal{B}\right)  $ and it is given as
\begin{equation}
I_{\phi}\left(  \mathcal{\rho},\mathcal{\sigma}\right)  =tr\theta_{\phi
}\left(  \log\theta_{\phi}-\log\left(  \rho\otimes\sigma\right)  \right)  .
\end{equation}
Besides this quantity describes an information gain in a quantum system
$\left(  \mathcal{A},\sigma\right)  $ via an entanglement $\phi_{*}$ with
another system ($\mathcal{B},\rho),$ it is naturally treated as a measure of
the strength of an entanglement, having zero the value only for completely
disentangled states (\ref{7.1}), corresponding to $\theta_{\phi}=\rho
\otimes\sigma$.

\begin{definition}
The maximal quantum mutual entropy for a fixed state $\sigma$
\begin{equation}
H_{\sigma}\left(  \mathcal{A}\right)  =\sup\{I_{\phi}\left(  \mathcal{A}%
,\mathcal{B}\right)  ;\phi_{*}\left(  I\right)  =\sigma\}
\end{equation}
is called q-entropy of the state $\sigma$. The differences
\begin{align*}
H_{\phi}\left(  \mathcal{B}|\mathcal{A}\right)   & =H_{\sigma}\left(
\mathcal{A}\right)  -I_{\phi}\left(  \mathcal{A},\mathcal{B}\right)  ,\\
D_{\phi}\left(  \mathcal{B}|\mathcal{A}\right)   & =S\left(  \mathcal{\sigma
}\right)  -I_{\phi}\left(  \mathcal{A},\mathcal{B}\right)
\end{align*}
are respectively called the q-conditional entropy on $\mathcal{B}$ with
respect to $\mathcal{A}$ and the degree of disentanglement for the compound
state $\phi$.
\end{definition}

$H_{\phi}\left(  \mathcal{B}|\mathcal{A}\right)  $ is obviously positive,
however $D_{\phi}\left(  \mathcal{B}|\mathcal{A}\right)  $ has the positive
maximal value $S\left(  \mathcal{\sigma}\right)  =\sup\left\{  D_{\phi}\left(
\mathcal{B}|\mathcal{A}\right)  ;\phi_{*}\left(  I\right)  =\sigma\right\}  $
and can achieve also a negative value
\begin{equation}
\inf\left\{  D_{\phi}\left(  \mathcal{B}|\mathcal{A}\right)  ;\phi_{*}\left(
I\right)  =\sigma\right\}  =S\left(  \mathcal{\sigma}\right)  -H_{\sigma
}\left(  \mathcal{A}\right)
\end{equation}
for the entangled states \cite{BO}, which is called the chaos degree
in\cite{IKO}.

Let us consider $\mathcal{G}$ as a Hilbert space describing a quantum input
system and $\mathcal{K}$ as its output Hilbert space. A quantum channel
$\Lambda^{*}$ sending each input state defined on $\mathcal{G}$ to an output
state defined on $\mathcal{K}.$ A deterministic quantum channel is given by a
linear isometry $\Upsilon$ $\mathrm{:}\mathcal{G}\rightarrow$%
\textrm{$\mathcal{K}$} with $\Upsilon^{\dagger}\Upsilon=I_{0}$ ($I_{0}$ is the
identify operator in $\mathcal{G}$) such that each input state vector $\eta
\in\mathcal{G}$, $\left\|  \eta\right\|  =1$ is transmitted into an output
state vector $\Upsilon\eta\in\mathcal{K}$, $\left\|  \Upsilon\eta\right\|
=1$. The mixtures $\rho=\sum_{n}p_{n}\omega_{n}$ of the pure input states
$\omega_{n}=\eta_{n}\eta_{n}^{\dagger}$ are sent into the mixtures
$\sigma=\sum_{n}p_{n}\sigma_{n}$ with pure states $\sigma_{n}=\Upsilon
\omega_{n}\Upsilon^{\dagger}$. A noisy quantum channel sends pure input states
$\omega$ into mixed ones $\sigma=\Lambda^{*}\omega$ given by the dual of the
following completely positive map $\Lambda$%
\begin{equation}
\Lambda\left(  A\right)  =\Upsilon^{\dagger}\left(  I_{1}\otimes A\right)
\Upsilon,\mathrm{\quad}A\in\mathrm{\mathcal{A}}%
\end{equation}
where $\Upsilon$ is a linear isometry from $\mathcal{G}$ to $\mathcal{F}%
_{1}\otimes$\textrm{$\mathcal{K}$}, $\Upsilon^{\dagger}\left(  I_{1}\otimes
I\right)  \Upsilon=I_{0}$, and $I_{1}$ is the identity operator in a separable
Hilbert space $\mathcal{F}_{1}$ representing the quantum noise. Each input
mixed state $\rho$ $\in B\left(  \mathcal{G}\right)  $ is transmitted into the
output state $\sigma=\Lambda^{*}\rho$ on $\mathcal{A}\subseteq B\left(
\mathcal{K}\right)  $, which is given by the density operator
\begin{equation}
\sigma=tr_{\mathcal{F}_{1}}\Upsilon\rho\Upsilon^{\dagger}\equiv\Lambda^{*}%
\rho\in\mathcal{A}_{*}.
\end{equation}

We apply the proceeding discussion of the entanglement to the above situation
containing a channel $\Lambda^{*}.$ For a given Schatten decomposition
$\rho=\sum_{n}p_{n}|n\rangle\langle n|\ $and the state $\sigma\equiv
\Lambda^{*}\rho,$we can construct three entangled states of the proceeding section:

(1) q-entanglement $\phi_{*}^{q}$ and q-compound state $\theta_{\phi}^{q}$ are
given as%

\begin{align*}
\phi_{*}^{q}(B)  & =\sum_{n,m}\left\langle n\mid B\mid m\right\rangle
tr_{\mathcal{F}}\kappa_{n}\kappa_{m}^{\dagger}\\
\theta_{\phi}^{q}  & =\sum_{m,n}|n\rangle\langle m|\otimes tr_{\mathcal{F}%
}\kappa_{n}\kappa_{m}^{\dagger}%
\end{align*}
with the marginals $\rho=\sum_{n}p_{n}|n\rangle\langle n|,$ $\sigma
\equiv\Lambda^{*}\rho=tr_{\mathcal{G}}\theta_{\phi}^{q}$ and $tr_{\mathcal{K}%
}\kappa_{n}\kappa_{m}^{\dagger}=p_{n}\omega_{n}\delta_{n}^{m}=\kappa
_{m}^{\dagger}\kappa_{n}$ for $\omega_{n}=\Lambda^{*}|n\rangle\langle n|.$ Let
$\mathcal{E}_{q}$ be the convex set of all completely positive maps $\phi^{q}$ .

(2) d-entanglement $\phi_{*}^{d}$ and d-compound state $\theta_{\phi}^{d}$ are
given as%

\begin{align*}
\phi_{*}^{d}(B)  & =\sum_{n}\left\langle n\mid B\mid n\right\rangle
tr_{\mathcal{F}}\kappa_{n}\kappa_{n}^{\dagger}\\
\theta_{\phi}^{d}  & =\sum_{n}|n\rangle\langle n|\otimes tr_{\mathcal{F}%
}\kappa_{n}\kappa_{n}^{\dagger}%
\end{align*}

\noindent with the same marginal conditions as (1). Let $\mathcal{E}_{d}$ be
the convex set of all completely positive maps $\phi^{d}.$

(3) c-entanglement $\phi_{*}^{c}$ and c-compound state $\theta_{\phi}^{c}$ are
same as those of (2) with commuting $\left\{  \omega_{n}\right\}  .$ Let
$\mathcal{E}_{c}$ be the convex set of all completely positive maps $\phi^{c}
$ .

Now, let us consider the entangled mutual entropy and the capacity of quantum
channel by means of the above three types of compound states.

\begin{definition}
The mutual entropy $I_{q}\left(  \rho,\Lambda^{*}\right)  $and the q-capacity
$C_{q}\left(  \Lambda^{*}\right)  $ for a quantum channel $\Lambda^{*}$are
defined by
\begin{align}
I_{q}\left(  \rho,\Lambda^{*}\right)   & =\sup\left\{  S(\theta_{\phi}%
^{q},\rho\otimes\Lambda^{*}\rho);\phi^{q}\in\mathcal{E}_{q}\right\}  ,\\
\;C_{q}\left(  \Lambda^{*}\right)   & =\sup\left\{  I_{q}\left(  \rho
,\Lambda^{*}\right)  ;\rho\right\}  .\nonumber
\end{align}
The d-mutual entropy, the d-capacity and the c-mutual entropy, the c-capacity
are defined as above using $\theta_{\phi}^{d}$ and $\theta_{\phi}^{c}$, respectively.
\end{definition}

Note that due to $\mathcal{E}_{c}\subseteq\mathcal{E}_{d}\subseteq
\mathcal{E}_{q},$ we have the inequalities
\begin{align*}
I_{q}\left(  \rho,\Lambda^{*}\right)   & \geq I_{d}\left(  \rho,\Lambda
^{*}\right)  \geq I_{c}\left(  \rho,\Lambda^{*}\right)  ,\;\\
C_{q}\left(  \Lambda^{*}\right)   & \geq C_{d}\left(  \Lambda^{*}\right)  \geq
C_{c}\left(  \Lambda^{*}\right)
\end{align*}
for a deterministic channel ($\Lambda^{*}=id$), the two lower mutual entropies
coincide with the von Neumann entropy:
\[
I_{d}\left(  \rho,id\right)  =-tr\rho\log\rho=I_{c}\left(  \rho,id\right)  .
\]
The capacity for such a channel is finite if $\mathcal{A}$ has a finite rank,
$C_{d}\left(  \Lambda^{*}\right)  \leq\dim\mathcal{K}$. On the other hand, the
q-mutual entropy can achieve the q-entropy
\[
I_{q}\left(  \rho,id\right)  =-2tr\rho\log\rho
\]
and its capacity is bounded by the dimension of the algebra $\mathcal{A}$,
$C_{q}\left(  \Lambda^{*}\right)  \leq\dim\mathcal{A}$ which doubles the
d-capacity dim$\mathcal{K}$ when $\mathcal{A}=B\left(  \mathcal{K}\right)  $.
These equalities will be related to the work on entropy by Voiculescu
\cite{Voi}.

\part{Information Genetics}

\section{Entropy Evolution Rate}

Genome sequence carries information as an order of four bases, and the
information is transmitted to m-RNA, which makes a protein as a sequence of
amino acids by a help of t-RNA.

In information theory, the concept of information has two aspects, one of
which expresses the amount of complexity of a whole system like a sequence
itself and another does the structure of the system(or message) such as the
rule stored in the order of sequence\cite{IKO}. From Shannon's philosophy, a
system has the larger complexity, the system carries the larger information,
from which the information of a whole system has been expressed by the
entropy. The structure of the system is studied in the field named ''coding
theory'', that is, how to code the messages is essential in communication of information.

Pioneering works for application of information theory to genome sequence were
done by Smith\cite{Smi}and Gatlin\cite{Gat}, since then few works have been
appeared along this line. In 1989 \cite{Ohy}, I introduced a measure
representing the difference of two genome or amino acid sequences, which is
called the entropy evolution rate and has been used to make phylogenetic
trees\cite{Ohy,MSO}. The coding theory was applied to the study of genome
sequences in order to examine the coding structure of several species\cite{OM}.

Let A and B be amino acid or base sequences. When they are considered to be
close each other, for instance, they specify an identical protein, we first
have to align these sequences by inserting a gap ''$*$\-'', whose arrangement
is called the alignment of sequences\cite{Sel,NW,OU}. As an example, take two
sequences $A$ and $B$ given as%

\[%
\begin{array}
[c]{l}%
A:\text{ a c b a c d }\\
B:\text{ a d b c a c b}%
\end{array}
\]
Then the aligned sequences become%

\[%
\begin{array}
[c]{l}%
A:\text{ a c b }*\text{ a c d }\\
B:\text{ a d b c a c b }%
\end{array}
.
\]

After the alignment, two sequences have the same length. Take two aligned
sequences $A$ and $B$ having the length n given by $A$=(a$_{1,}$ a$_{2,}%
\cdot\cdot\cdot,$ a$_{n}),$ $B$=(b$_{1,}$ b$_{2,}$ $\cdot\cdot\cdot,$ b$_{n}$
), where a$_{i,}$ b$_{i}$ are the gap $*$ or an amino acid for an amino acid
sequence or a base for a base sequence. There are 21 events (20 amino acids
and $*$) in an amino acid sequence and 5 events (4 bases and $*$) in a base
sequence. Therefore, in an aligned sequence, the occurrence probability of
each amino acid (resp. base) is associated, and it is denoted by $p_{k}$ for
k-th amino acid (resp. base), where $0\leq k\leq20 $ (resp. $0\leq k\leq4)$
and ''0'' corresponds to the gap. Then the entropy (information) carried by
the amino acid (resp. base) sequence $A$ is defined as%

\[
S(A)(\text{or }S(p)\text{)}=-\sum_{k}p_{k}\log p_{k}
\]
where $p$ denotes the probability distribution ($p_{k}$). Similarly, there
exists the event system $(B,q\equiv(q_{k})$ ) for the amino acid (or base)
sequence $B,$ and its entropy is denoted $S(B)$ or $S(q).$ Through the
alignment, we can find the correspondence between the amino acid (resp. base)
of $A$ and that of $B,$ which enables to make the compound event system
($A\times B,$ $r)$ of $A$ and $B.$ Here $r$ is the joint probability
distribution between $A$ and $B,$ so that it satisfies $\sum_{k}r_{jk}=p_{j} $
and $\sum_{k}r_{jk}=q_{k}.$

The most important information measure in Shannon's communication theory is
the mutual entropy (information) expressing the amount of information
transmitted from ($A,p)$ to ($B,q),$ which is defined as follows:%

\[
I(A,B)=\sum_{j,k}r_{jk}\log\frac{r_{jk}}{p_{j}q_{k}}.
\]
Using the entropy and the mutual entropy, an quantity measuring the similarity
between $A$ and $B$ was introduced as%

\[
r(A,B)=\frac{1}{2}\left\{  \frac{I(A,B)}{S(A)}+\frac{I(A,B)}{S(B)}\right\}  ,
\]
which was called the symmetrized entropy ratio or the entropy evolution rate
in \cite{Ohy}and it takes the value 0 when $A$ and $B$ are completely
different and 1 when they are identical. The minus of this rate from 1
indicates the difference between $A$ and $B$. We here call it the entropy
evolution rate, and it is denoted by $\rho(A,$ $B)$ :%

\[
\rho(A,B)=1-r(A,B).
\]
Using this rate, we can construct a genetic matrix and write a phylogenetic
tree of species\cite{Ohy,MSO}. Note that a similar measure providing the
difference between $A$ and $B$ can be defined as%

\[
\rho^{\prime}(A,B)=1-\frac{I(A,B)}{S(A)+S(B)-I(A,B)},
\]
but this dose not have a precise meaning from the information theoretical
point of view.

An application of this rate to the variation of HIV virus for six patients
reported by \cite{WZB,HZS,MHM,JGK}is discussed in \cite{SMO}.

\section{Code Structure of Genes}

When we send an information (a series of messages), we have to process the
messages in proper forms so as to correctly and quickly send the information
to a receiver. It is the coding theory that teaches us how to process the
messages properly. There are many ways to encode the messages in communication
processes. We shall explain some of such codings and their use to the study of
genome sequences.

Let $i=(i_{1},i_{2},$ $\cdot\cdot\cdot,i_{k})$ be a properly processed
information sequence. In order to send the symbol $i$ to a receiver correctly,
that is, to avoid some noise and loss in the course of information
transmission, we have to add some redundancy (parity check symbol)
$p=(p_{1},p_{2},\cdot\cdot\cdot,p_{n-k})$ to the information symbol $i.$ This
redundancy $p$ detects or corrects the errors in the communication process.
The whole code-word now becomes%

\[
x=(i_{1},i_{2},\cdot\cdot\cdot,i_{k},p_{1},p_{2},\cdot\cdot\cdot,p_{n-k}).
\]
The above $x$ is called a systematic code, and to make the systematic code $x
$ from the information symbol $i$ is called a coding. A coding is realized by
a Galois group $GF(q)$ with a primary number $q$ and a certain parity check
$p.$ When the relation between $i$ and $p$ is linear, the code so obtained is
called a linear code. Among the linear codes, there are the block code such as
cyclic code and BCH code and the convolutional code such as self-orthogonal
code and Iwadare code. Each code has its own parity check correcting the error
such as random error, burst error and bite error. We do not go into the
details of the coding theory here, but we explain how to use the coding
technique to examine the code structure of genome sequences.

When we like to know the code structure of a species, an organism, a special
part of a genome sequence indicating a protein or a set of these objects, we
rewrite a base sequence of an object into the sequence of the symbols of
$GF(2^{2})$ because we have four bases, and we apply several coding methods to
the symbol sequence and get the coded symbol sequence (systematic code), then
we write it back the coded base sequence. This process is written as follows:
\[%
\begin{array}
[c]{l}%
\text{ Base sequence }A\Longrightarrow\text{Symbol sequence }A_{s}\\
\Longrightarrow\text{Coded symbol sequence }A_{s}^{C}\Longrightarrow\text{
Coded base sequence }A^{C}%
\end{array}
\]
In order to know the common code structure of the sequences $A_{1,}A_{2,}%
\cdot\cdot\cdot,A_{n},$ we use the following index obtained from the entropy
evolution rate and a coding $C$ applied to the sequences:%

\[
D_{C}=\frac{\left\{  \sum_{i=1}^{n-1}\sum_{j=i+1}^{n}\left|  \rho(A_{i}%
,A_{j})-\rho\left(  A_{i}^{C},A_{j}^{C}\right)  \right|  \right\}
}{_{\text{n}}\text{C}_{\text{2}}},
\]
where $_{\text{n}}$C$_{\text{2}}$ is the combination 2 out of n, that is,
$_{\text{n}}$C$_{\text{2}}=\frac{n(n-1)}{2}$ and $A_{i}$ is an amino acid
sequence or a base sequence$.$ Note that when $A_{i}$ is originally an amino
acid sequence, we first translate it the corresponding base sequence and take
the above procedure, then we convert the coded base sequence to the coded
amino acid sequence. If this index $D_{C}$ is close to 0, then a common code
structure of the group $\left\{  A_{1,}A_{2,}\cdot\cdot\cdot,A_{n}\right\}  $
is close to the structure of the code $C$ used.

We studied the code structure of Vertebrate, Onco virus and HIV virus by means
of the structure index $D_{C}.$ We used some parts of the base sequence for
each organisms; MDH, LDH, hemoglobin $\alpha,$ $\beta$ for Vertebrate; pol,
env, gag for Onco and HIV virus. Then we obtained the following results:

(1) Vertebrate has a similar code structure of the convolutional code with
high ability correcting the burst errors like the codes named UI, ZI, and the
code structure of hemoglobin $\alpha$ is closest to that of the artificial codes.

(2) Onco virus has a similar code structure of the cyclic code with the burst
error correction (C2) or the self-orthogonal code (TB,VD), so that it does not
have so high ability correcting the errors.

(3) HIV virus has a similar code structure of the cyclic code (C1) or the
self-orthogonal code with the random error correction (TA) , so that the
ability correcting the errors is low.

(4) In Onco and HIV virus, the pol protein has the closest code structure of
the artificial codes.

\smallskip

The structure index is applied to the study of the variation and the condition
of the patients having the HIV\ infection in \cite{TSO}.

\end{document}